\documentclass[amsmath,amssymb,aps,prb,reprint,superscriptaddress,nobibnotes]{revtex4-2}

\usepackage{amsmath}
\usepackage{amsfonts}
\usepackage{amssymb}
\usepackage{graphicx}
\usepackage{upgreek}
\usepackage[dvipsnames]{xcolor}
\usepackage{braket}
\usepackage{verbatim}
\usepackage{bm}

\makeatletter 
\renewcommand\@biblabel[1]{#1.} 
\makeatother

\definecolor{darkred}{rgb}{0.5, 0, 0}
\definecolor{darkgreen}{rgb}{0, 0.5, 0}
\definecolor{darkblue}{rgb}{0.1, 0.1, 0.7}

\usepackage[bookmarks=true,
            bookmarksnumbered=false,
            bookmarksopen=true,
            breaklinks=true,
            pdfborder={0 0 0},
            final=true,
            backref=none,
            colorlinks=true,
            linkcolor=darkblue,
            menucolor=darkblue,
            citecolor=darkblue,
            urlcolor=darkblue]{hyperref}

\newcommand{\micro}{${\upmu}$}

\newcommand{\upsub}[1]{_{\mathrm{#1}}}
\newcommand{\um}{$\,$\micro m}
\newcommand{\uev}{$\,$\micro eV}

\begin{document}

\title{Flat-band compactons in a two-dimensional driven-dissipative Lieb lattice}

\author{Seth Lovett}
\author{Paul M. Walker}
\email{p.m.walker@sheffield.ac.uk}
\author{Anthony Ellul}
\affiliation{School of Mathematical and Physical Sciences, University of Sheffield, S3 7RH, Sheffield, UK}
\author{Edmund Clarke}
\affiliation{EPSRC National Epitaxy Facility, University of Sheffield, Sheffield S3 7HQ, UK}
\author{Maurice S. Skolnick}
\author{Dmitry N. Krizhanovskii}
\affiliation{School of Mathematical and Physical Sciences, University of Sheffield, S3 7RH, Sheffield, UK}

\begin{abstract}
We experimentally study the effect of inter-particle interactions on the flat-band states of a two-dimensional Lieb lattice with drive and dissipation. Exploiting the giant nonlinear interactions of exciton polaritons we observe compactly localised solitons (compactons) embedded within the continuum of propagating state bands - a form of nonlinear bound state in the continuum (BIC). The driven-dissipative nature of the system leads to a sudden self-localisation into the compacton state above a threshold power. The experimental results agree well with numerical simulations. These results have implications for the physics of interacting quantum particles in flat-band systems and for generation of quantum-correlated light and spatially multiplexed coherent information transport.
\end{abstract}

\maketitle



Flat band states occur in many natural and artificial crystal structures~\cite{Leykam2018,Leykam2018APX}. They arise when properties of the lattice lead to a band where the energy does not depend on momentum. This quenching of kinetic energy leaves dissipation and interactions as the dominant energy scales resulting in a host of important many-body effects including fractional quantum Hall phases~\cite{wang2011fractional}, high temperature superconductivity~\cite{Peotta2015}, Wigner crystallization~\cite{wu2007flat}, negative magnetism~\cite{yin2019negative} and disorder induced topological phase transitions~\cite{chen2017disorder}. A particularly important testbed for studying flat-band physics is the two-dimensional (2D) Lieb lattice, where the chiral symmetry of the unit cell leads to a flat band topologically protected against disorder in the site-to-site coupling strength~\cite{Leykam2018APX}. 

A key property of flat bands is that they support compact localised states (CLS). These are maximally localized superpositions of momentum states from over the whole Brillouin Zone, enabled by the massive degeneracy~\cite{Leykam2018}. They have the distinctive property that the localisation is even stronger than exponential, with the wavefunction strictly zero outside some spatial region. In the 2D Lieb lattice the flat band states are embedded within dispersive bands~\cite{GuzmanSilva2014}. CLS, however, do not couple to the propagating continuum and remain fully localised. They may therefore be regarded as a kind of bound-state in the Continuum (BIC), a topic which has recently excited intensive research activity in the photonics community~\cite{Hsu2016,Azzam2021}. Compactly localised solitons, or compactons, which remain compactly localised even in the presence of interactions, have been predicted in a variety of lattices~\cite{Yulin2013,Gligoric2016,Zegadlo2017,Real2018}. 2D flat bands have been the subject of intense experimental investigation in the non-interacting case~\cite{GuzmanSilva2014, Mukherjee2015, Vicencio2015,Kajiwara2016,klembt2017polariton,whittaker2018exciton,harder2020exciton}, and theoretical investigation of the interacting case~\cite{Leykam2012,Vicencio2013,Lazarides2017,Belicev2017,Leykam2018APX,Real2018}. Despite its importance, however, the interacting regime of 2D flat band states could not be directly experimentally accessed until now.

Here we study flat band states in a photonic lattice with strong interactions and find compactons that are part of a previously unobserved class originating at the flat band at zero power~\cite{Real2018}, even when embedded within dispersive bands and thus distinct from gap solitons~\cite{Alexander2006,goblot2019nonlinear}. We further find that the effect of drive and dissipation can cause sudden self-localisation of particles into these solitonic states. The results open up new regimes for experimental study of interacting quantum particles in flat bands and simulation of novel magnetic phases~\cite{Leykam2018}, and new practical directions for generation of quantum correlated states~\cite{Biondi2015,Casteels2016}, classical all-optical logic gates~\cite{Real2017}, and high density spatially multiplexed information transport~\cite{Vicencio2015,Vicencio2014}.

To reach this regime we exploit the favorable properties of exciton-polaritons~\cite{schneider2016exciton,Amo2016}. These are hybrid particles which are part photon and part quantum-well exciton. A large variety of potential landscapes can be imposed through photonic confinement, while the excitonic part allows giant interparticle interactions. Photon tunnelling through cavity mirrors naturally provides direct imaging of the internal cavity wavefunction and allows study of driven-dissipative physics where external pumps balance losses and can drive systems to a desired equilibrium state~\cite{Jamadi2022,heras2024}. This makes polaritonic lattices an excellent platform for studying the interplay of novel single-particle bandstructure and nonlinear interactions.

So far, lasing of flat band states in polariton Lieb lattices was studied~\cite{Baboux2016,klembt2017polariton,whittaker2018exciton,harder2020exciton}, though these works did not address individual CLS with coherent external drive or consider the Kerr-nonlinear-like polariton interactions. Coherent excitation of CLS in the linear (non-interacting) regime was performed in waveguide~\cite{Vicencio2015,Mukherjee2015} and micropillar cavity~\cite{harder2020exciton} geometries. A one-dimensional (1D) analogy to the Lieb lattice was studied experimentally~\cite{Baboux2016,goblot2019nonlinear}, but is qualitatively different both due to reduced dimensionality and because in that case the flat band is separated from the dispersive bands by an energy gap. The nonlinear states are explained as gap solitons~\cite{goblot2019nonlinear} which are spatially localised due to the absence of propagating states in the gap rather than due to the lattice topology as in our case. Similarly, a recently studied electrical diamond lattice was 1D with gapped flat band~\cite{ChaseMayoral2024}. Apart from the localisation by topology that we are primarily concerned with, reconfigurable localisation using precisely tuned resonant driving has also been studied experimentally~\cite{Jamadi2022} and theoretically~\cite{heras2024}.

\begin{figure} 
    \centering
    \includegraphics[width=\columnwidth]{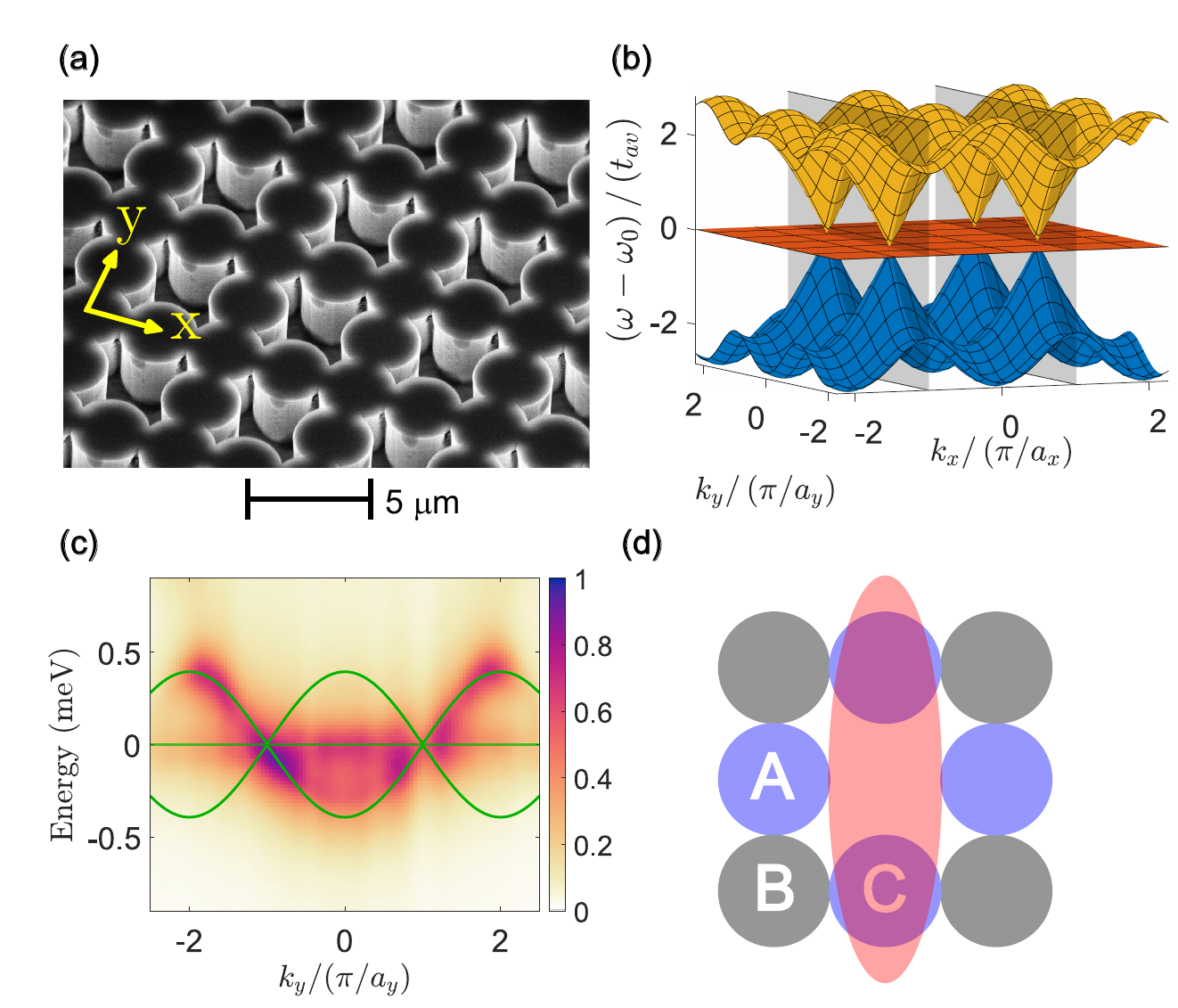}
    \caption{(a) Scanning electron microscope image of part of the Lieb lattice micropillar array. (b) Lieb lattice bandstructure calculated using a tight binding model. $t_{av}=\left(t\upsub{TE}+t\upsub{TM}\right)/2$. The flat band is sandwiched between dispersive bands with no band gap. (c) Angle-resolved photoluminescence spectrum showing a section through the real bandstructure of the experimental device. The image is an average over two sections at fixed $k_x = \pm\pi/a_x$ (shown by gray planes in panel (b)). Solid lines show the expected bands calculated using the tight binding model with parameters $t\upsub{TE}=0.20$~meV and $t\upsub{TM}=0.22$~meV. (d) Schematic showing the pump spot profile used in our work (red ellipse) superimposed over a single plaquette of our Lieb lattice (circles). Letters A, B, C indicate the three sites in one unit cell of the lattice. The circles shaded blue indicate the sites with non-zero population for an ideal CLS in a Lieb lattice~\cite{Vicencio2015}.}
    \label{fig:schematic}
\end{figure}

\section*{Results}
In this work we study a 2D Lieb lattice of polariton micropillar resonators. Figure~\ref{fig:schematic}(a) shows a scanning electron microscope image of the lattice we used, which was fabricated alongside previously reported devices~\cite{whittaker2018exciton}. Details of the fabrication are given in the \hyperref[sec:methods]{Methods}. Fig.~\ref{fig:schematic}(d) shows a schematic of one plaquette - the smallest unit of the lattice capable of supporting a CLS~\cite{Baboux2016}. Here it consists of a loop of 8 pillars with only the 4 shaded in blue having non-zero intensity for an ideal CLS~\cite{Vicencio2015}. The underlying Lieb lattice has three sites per unit cell, which are indicated by the letters A,B,C for one particular unit cell. 

Figure~\ref{fig:schematic}(b) shows the expected bandstructure for $y$-polarised photons calculated using a tight binding model (see Supplementary Discussion 1). There are three bands per polarisation, as expected for a three site unit cell, and two decoupled polarisations. One of the bands is flat while the other two mirror each other above and below the flat band. The two dispersive bands have finite slope and so support photons propagating with non-zero velocity in the $x$ and $y$ directions.
Figure~\ref{fig:schematic}(c) shows a slice of the single-particle bandstructure of the real lattice measured using angle-resolved photoluminescence (PL) spectroscopy, consistent with the results in Ref.~\cite{whittaker2018exciton}.
The solid lines on the figure show the expected energies from the tight binding model using hopping rates of $t\upsub{TE}=0.20$~meV and $t\upsub{TM}=0.22$~meV for light polarised perpendicular and parallel to the direction between sites respectively. The average linewidth in PL is 0.3 meV.

To study the response of the system to external driving we resonantly excited a spatially small region of the lattice using an elliptical shaped laser spot as illustrated in the schematic Fig.~\ref{fig:schematic}(d). We began with the laser overlapping two C sites of the lattice. We monitored the occupancy of all the sites in the lattice by collecting the light coming from the opposite side of the cavity from where the laser was incident (transmission geometry) and imaging this onto a CCD camera.

\begin{figure*}
    \centering
    \includegraphics[width=\textwidth]{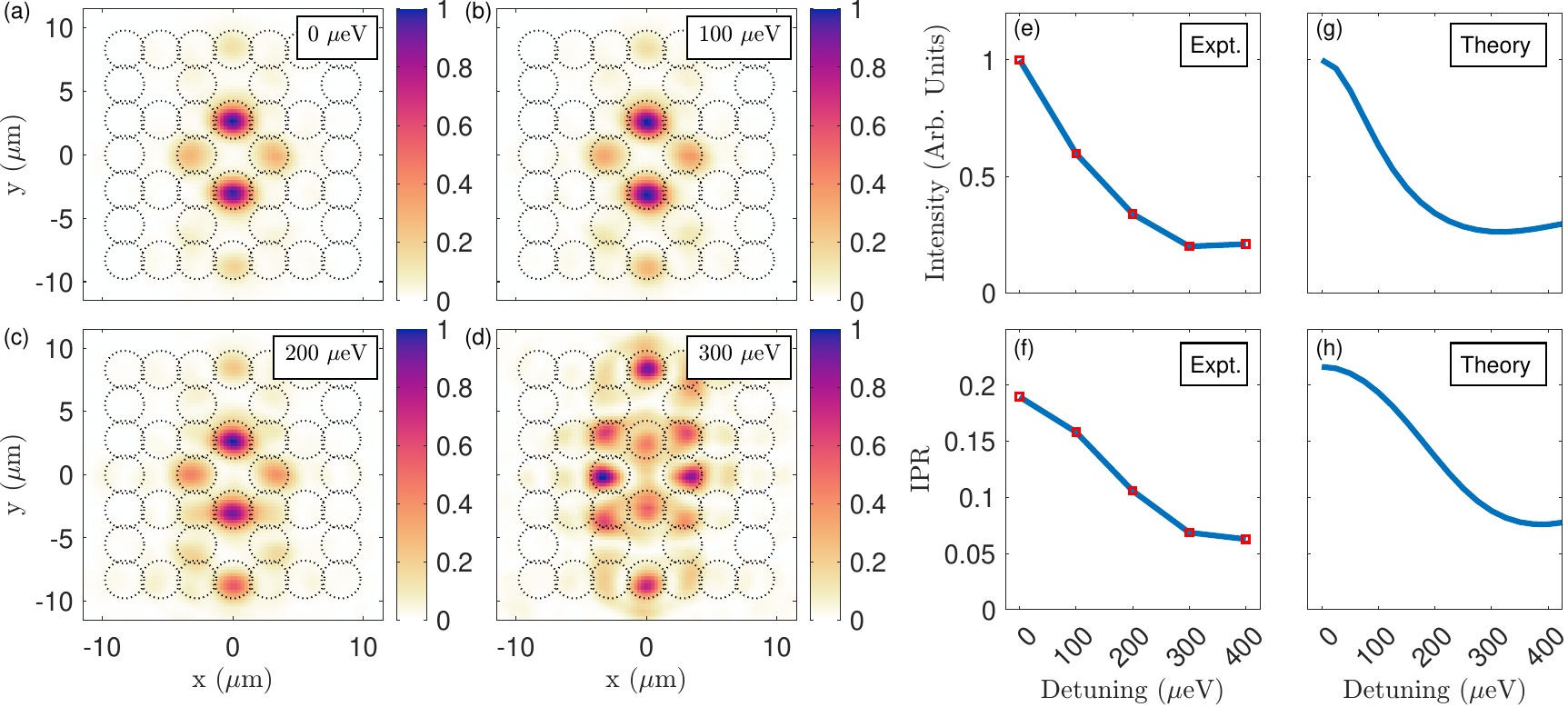}
    \caption{(a-d) Spatial emission profile when pumping two C sites at fixed power of 6mW and varying energy detuning $\Delta$E of the laser relative to the flatband. The values of $\Delta$E are shown in the top right of each panel. (e) Experimental total intensity of all sites on the lattice $\Delta$E. (f) Experimental IPR vs. $\Delta$E calculated using Eqn.~\eqref{eq:IPR} (see Supplementary Discussion 4). (g) Total intensity vs. pump strength from numerical model. (h) IPR vs. pump strength from numerical model.} 
    \label{fig:linear}
\end{figure*}

We first consider the non-interacting properties of the lattice. We tuned the laser frequency relative to the flat band and imaged the distribution of polaritons among the lattice sites. We used an incident laser power of 6 mW, sufficiently low that interactions were negligible. Figure~\ref{fig:linear}(a) shows the measured occupancy of the sites for zero detuning $\Delta E$ of the laser energy from the flat band.
The two C sites that are overlapped by the laser spot have the highest intensity. The two A sites on the same plaquette have the next highest intensity, while the B sites at the corners have low occupation. There is also some intensity on other pillars surrounding the directly excited plaquette.

Occupation of the A and C sites on the pumped plaquette with zero emission from the B sites is the expected profile for a CLS (see schematic in Fig.~\ref{fig:schematic}d and Supplementary Discussion 1C). The zero emission on the B site results from destructive interference between the A and C site populations, which have opposite phase~\cite{whittaker2018exciton}, and is a key characteristic of CLS. In our experiment we see nearly but not quite zero B site emission. We also see brighter emission from the C sites compared to the A sites. This was also observed in Ref.~\cite{Vicencio2015} and arises because we do not excite the system with the ideal CLS intensity and phase profile, but rather only excite the C sites. This leads to population of dispersive band states which interfere with the CLS as well as propagating out into the lattice. This propagation also explains the intensity on sites away from the pumped plaquette. In Fig.~\ref{fig:linear}(b-d) we show the result of gradually increasing the detuning $\Delta E$ of the laser energy above the flat band energy. The emission increasingly comes from sites away from the pumped plaquette and the population of the B sites eventually becomes comparable to that of the A and C sites. The total intensity summed over all the sites is shown in Fig.~\ref{fig:linear}(e). As the detuning is increased the overall intensity decreases. This is expected because as the energy is tuned out of the highly degenerate flat band into the dispersive bands, which have lower density of states, the incident laser couples into the cavity less efficiently.

The degree to which the population is spatially localised can be quantified using the inverse participation ratio~\cite{Vicencio2015,Jamadi2022} given by

\begin{equation}\label{eq:IPR}
     \mathrm{IPR}=\frac{\sum_j I_j^2}{(\sum_j I_j)^2}.
\end{equation}

\begin{figure*}
    \centering
    \includegraphics[width=\textwidth]{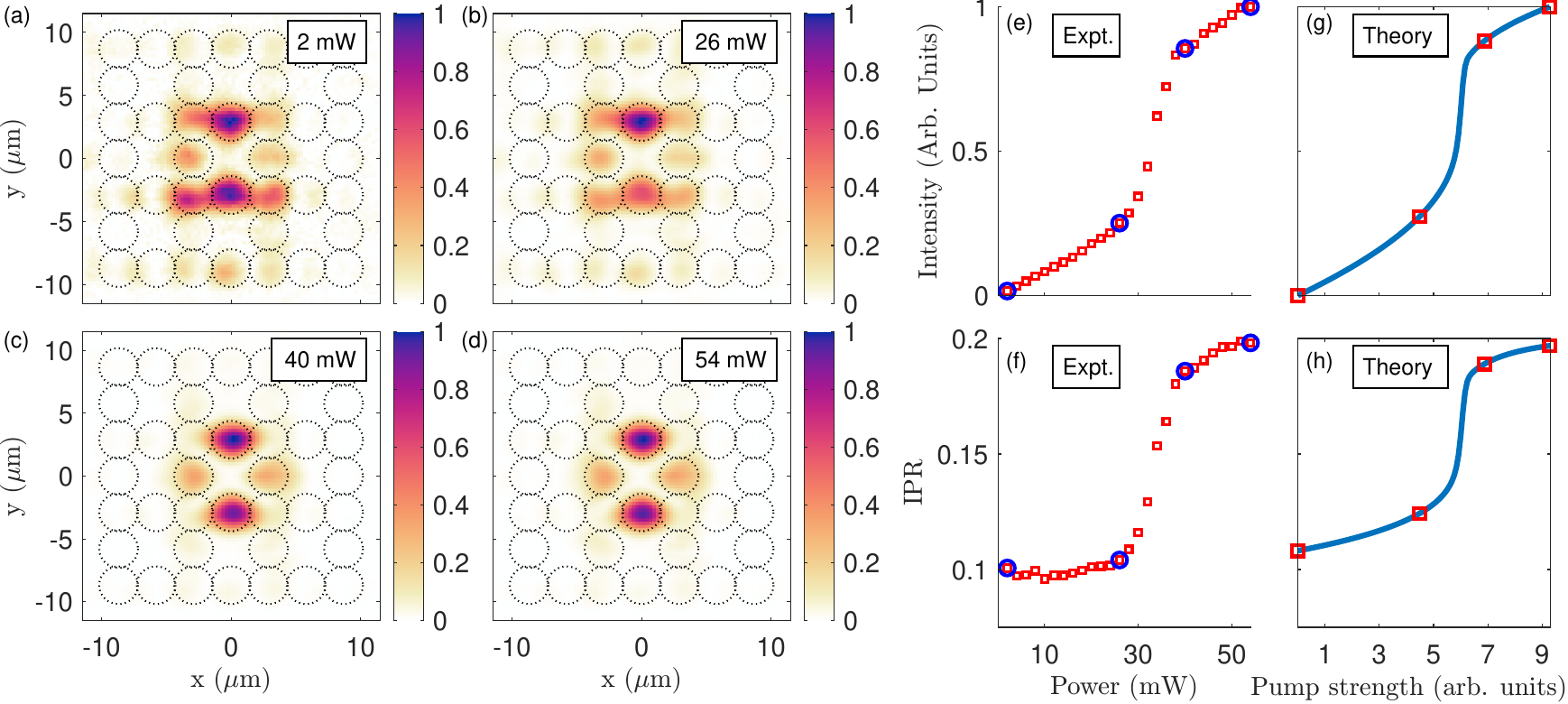}
    \caption{Power dependence. (a-d) Experimental spatial emission profile when pumping two C sites for four powers and a fixed energy detuning of $\Delta E$ = +250\uev{}. Intensity is normalised to the peak in all images. (e) Experimental total intensity summed over all sites vs. power. (f) IPR vs. power. The powers corresponding to panels (a-d) are indicated by blue circles. (g) Total intensity vs. pump strength from numerical model. (h) IPR vs. pump strength from numerical model. The numerically modeled spatial profiles corresponding to the red squares are shown in Fig.~\ref{fig:theory_results}.}
    \label{fig:nonlinear}
\end{figure*}

Here $I_j$ is the intensity on site $j$ (see Supplementary Discussion 4). The IPR takes the value 1 if all the intensity is on a single site and $1/N$ for delocalisation over $N$ sites. Thus higher values of IPR mean tighter spatial localisation. For an ideal CLS we expect an IPR $=0.25$~\cite{Vicencio2015}. Figure~\ref{fig:linear}(f) shows how the IPR varies as we tune the laser. At zero detuning (resonant with the flat band) the IPR is 0.19. This indicates the population is not as localised as would be expected for a CLS, consistent with the additional weak excitation of the dispersive states. With increasing detuning the IPR drops even lower as the populations spreads further away from the pumped plaquette, reaching 0.06 when $\Delta E=0.4$ meV.

To measure the effect of the nonlinear interactions we fixed the laser frequency at $\Delta E=0.25$ meV above the flat band and gradually increased the intensity. The spatial distributions for four powers are shown in Fig.~\ref{fig:nonlinear}(a-d). The total intensity and IPR vs. incident power are shown in Fig.~\ref{fig:nonlinear}(e) and Fig.~\ref{fig:nonlinear}(f) respectively. At the lowest power the A, B and C sites are all populated, along with sites far from the pumped plaquette, and the IPR is $\sim 0.1$ indicating significant population of propagating modes. For powers below $\sim$30 mW the pattern and IPR remain qualitatively unchanged while the intensity of the two pumped sites increases linearly with incident power. Above $\sim$30 mW the IPR rapidly increases to $\sim 0.2$ and the pattern changes to resemble that in Fig.~\ref{fig:linear}(a) with close to zero population on the B sites of the pumped plaquette or outside the plaquette. At the same time there is a rapid increase in total intensity. Above $\sim$40 mW the pattern and IPR remain approximately stable and the intensity increases linearly again. We thus observe a sudden jump between two regimes. Below threshold there is significant population in the dispersive band propagating states. Above threshold the system jumps to the CLS-like state with a small residual population in dispersive states, and then remains stable.


We now compare our experimental results with numerical solutions for the expected steady state of the system. We use an established model for polariton microcavity lattices and other coupled driven-dissipative resonators~\cite{heras2024,Real2018} (see \hyperref[sec:methods]{Methods} and Supplementary Discussion 1 for more details). We first consider the non-interacting case. Figures~\ref{fig:linear}(g) and \ref{fig:linear}(h) show the total intensity and the IPR respectively for varying detuning $\Delta E$ and a loss rate $\gamma$ = 0.25 meV, close to the PL linewidth.
The semi-quantitative agreement with the experiment (Figs.~\ref{fig:linear}(e,f)) indicates that the model accurately captures the essential physics of the balance of drive, loss and photon hopping through the lattice.

\begin{figure}
\includegraphics[width=\columnwidth]{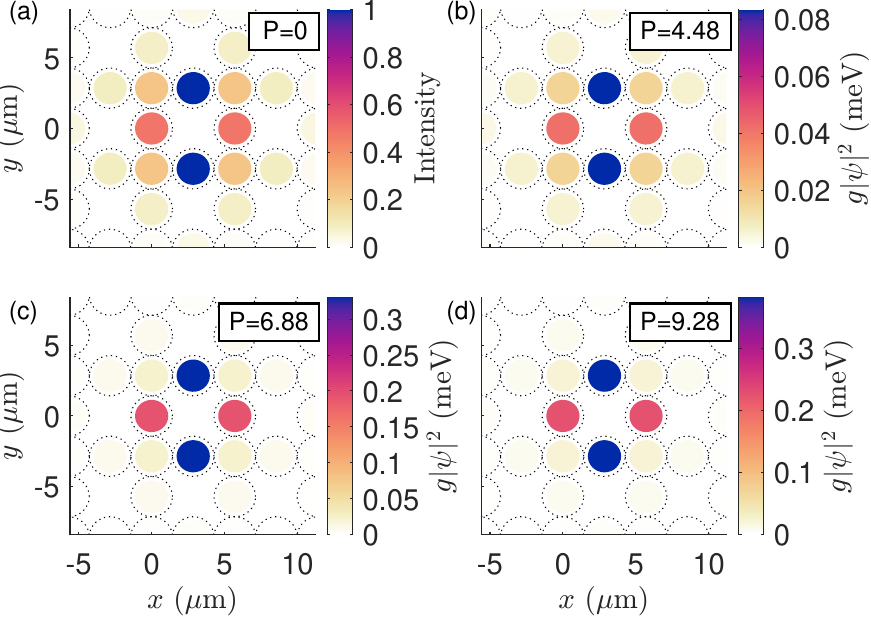}
\caption{Numerically calculated spatial distribution of intensity. (a-d) show distributions with increasing the pump strength $P$ as labelled in the top right. The pump strengths correspond to the red squares in Fig.~\ref{fig:nonlinear}(g,h). For (b-d) the color bars give the intensity multiplied by the interaction constant $g$ for linearly polarised polaritons. This is the size of the nonlinear potential energy on each pillar. For panel (a) the simulation is for negligible interactions so the color bar gives intensity normalised to the peak.}
\label{fig:theory_results}
\end{figure}

We now consider the results of the numerical modeling for different pumps powers and detuning $\Delta E=0.25$ meV, corresponding to our experimental observations in Fig.~\ref{fig:nonlinear}(a-f). Figures~\ref{fig:nonlinear}(g) and \ref{fig:nonlinear}(h) show the total intensity and IPR, both next to the corresponding experimental data for comparison. Figures.~\ref{fig:theory_results}(a-d) show the spatial distribution of intensity for the powers indicated by red squares in Figs~\ref{fig:nonlinear}(g,h). Again, semi-quantitative agreement of the IPR and total intensity with the experimental values are achieved. The spatial patterns show an abrupt switch to being highly concentrated on the A and C sites of the pumped plaquette above threshold, as also seen in the experiment. Although not shown in this figure we note that above threshold there is a $\pi$ phase difference between the fields on the A and C sites as expected for CLS~\cite{whittaker2018exciton} (see Supplementary Discussion 1C). Overall, the good agreement between experiment and simulations indicates that the effect we measure experimentally is due to the polariton interactions and their interplay with pumping, loss, and the lattice bandstructure.


The response to several other pump spatial distributions was also investigated, with detailed discussion in Supplementary Discussion 2. We experimentally and numerically investigated the case considered in Refs.~\cite{Jamadi2022,heras2024} where localisation in the non-interacting regime was observed due to pumping multiple adjacent sites. When pumping three pillars in a column (sites BAB with strongest pump on the A site) we found at low power the results were consistent with those works. The localisation of the field depends on the details of the pump intensity and phase on different sites. Increasing the pump power may then either localise of de-localise the field depending on the initial degree of localisation.

We also numerically calculated the response to a wide range of other pump conditions. Of particular interest is the case where the pump has the profile of an ideal CLS (Supplementary Fig. S11). This is the case where at low power there is the highest possible overlap with flat band states and the lowest overlap with propagating states. In this case we found that the IPR was close to 0.25 for all pump powers, but there is a strong jump in total intensity at a threshold power. The intensity distribution is initially a highly localised CLS and the nonlinearity preserves this structure. Also interesting is the case where a single B site is pumped (Supplementary Fig. S9). This is the opposite case - at low power there is the minimum overlap with flat band states and only propagating states are excited. Here the intensity distribution becomes increasingly delocalised with increasing power while the total intensity increases sub-linearly.

Overall, we find in both experiment and simulations that when the pump has sufficient overlap with the non-interacting regime CLS then the nonlinearity serves to drive the field closer to a nonlinear CLS, or compacton, state. The localisation occurs rapidly around a threshold power and is accompanied by a sudden jump in total intensity. For other pumps the nonlinearity may either localise or de-localise the field depending on the precise pump conditions that determine which dispersive and flat band states are populated at low power. 


\section*{Discussion}

A qualitative explanation for our observations is provided by the chiral symmetry of the Lieb lattice which, for CLS, forces destructive interference and zero intensity on B sites~\cite{whittaker2018exciton}. This isolates CLS from the rest of the lattice, with this geometric frustration leading to the flat band. An ideal compacton has the same field as the CLS and so the associated nonlinear potential does not break this symmetry, so a compacton is still isolated from the rest of the lattice. Thus, if a CLS is efficiently excited then a compacton will emerge and remain present as the power is increased. In Ref.~\cite{Real2018} Real and Vicencio demonstrate analytically that perfect compacton solutions are indeed exact stationary solutions in the nonlinear regime and bifurcate from the linear regime CLS at zero threshold power. We experimentally and numerically observe such a state that retains the profile of a CLS even at elevated powers. Qualitatively similar effects were also studied for various 1D chains~\cite{Yulin2013,Gligoric2016,Zegadlo2017} making this an example of a broad class of compactly localised solitons.

A key additional feature of our present system is that it has dissipation and external drive. We pump at a frequency above the flat band and there is a sudden jump in intensity at a threshold power. Such intensity jumps are well understood as the result of polariton states blueshifting into resonance with an external pump with increasing power. They have been observed for single isolated micropillars~\cite{Boulier2014} and planar cavities with quasi-plane-wave pumping~\cite{Baas2004}. Our observation of them here is evidence that the states we observe have a frequency which blueshifts with increasing power while they retain the localised CLS profile, noting that when the pump does not overlap the CLS profile the jumps do not occur (see Supplementary Discussion 2). Such a state is exactly the compacton predicted in Ref.~\cite{Real2018}.

The jump in IPR accompanying the intensity jump may be explained as follows. In general an arbitrary pump profile will excite both compacton and propagating states. However, the blueshift for the more spread-out propagating states is lower (Supplementary Discussion 2) so the compacton will jump to high intensity at a lower threshold. The result is the observed preferential population of the compacton above threshold. On the other hand, if the low-power overlap of pump and CLS is poor then other effects like those from Refs.~\cite{Jamadi2022} and~\cite{heras2024} will dominate, as we saw when we excited a column of pillars (Supplementary Discussion 2). For completeness we also compare our results with a number of studies on related but different systems in Supplementary Discussion 3.

To summarise, we experimentally study the effect of nonlinear interactions on the flat-band states of a two-dimensional Lieb lattice and observe compacton states embedded within dispersive bands. To our knowledge this is the first experimental observation of this class of discrete solitons~\cite{Real2018}. The driven-dissipative nature of the system leads to sudden preferential occupation of the compacton, which appears as a kind of self-localisation. The results are likely to be important for the physics of interacting quantum particles in driven-dissipative systems. From an applications perspective, nonlinear flat band states have been proposed as a route towards classical all-optical logic gates~\cite{Real2017} and generation of quantum-correlated multi-photon states~\cite{Biondi2015,Casteels2016} where drive and dissipation and the quenching of kinetic energy play key roles. Compactons in 2D lattices have also been proposed for diffraction-free high density spatially multiplexed information transport~\cite{Vicencio2015,Vicencio2014}, where the nonlinear reinforcement of localisation is likely to be beneficial.


\section*{Methods}\label{sec:methods}
\subsection*{Experiment}
The lattice (see Fig.~\ref{fig:schematic}(a)) was made by first growing a planar multilayer semiconductor Fabry-Perot cavity that confines light in the $z$-direction. It consists of two GaAs/Al$_{0.85}$Ga$_{0.15}$As Bragg mirrors sandwiching a $\lambda$ thickness GaAs cavity containing 3x 10 nm In$_{0.04}$Ga$_{0.96}$As quantum wells, one on each antinode of the electric field. This cavity was then patterned into a 2D Lieb lattice by electron beam lithography and inductively coupled plasma reactive ion etching. The depth of the etch reached to a few DBR pairs above the cavity layer. The circular pillars of diameter 3\um{} (see Fig.~\ref{fig:schematic}(a)) are the lattice sites where the photons are laterally confined. A slight overlap of the pillars allows hopping of the photons from site to site and hybridisation of the individual pillar eigenstates into bands~\cite{Amo2016}. The pillar center-to-center separation is 2.85\um{} giving a lattice period of 5.7\um{}. The lattice is immediately adjacent to the one used in previously reported works~\cite{whittaker2018exciton}, on the same 5 mm x 5 mm chip, and was patterned and etched in the same process run. The flat band energy determined from PL is 6.7 meV below the quantum well exciton energy of 1.476436 eV. The vacuum Rabi splitting, which characterises the photon-exciton coupling rate, is 4.7 meV~\cite{whittaker2018exciton}. From these we estimate that at the frequency of the flat band the polaritons have 89\% photonic content and 11\% excitonic content.

\subsection*{Numerical model}
The model consists of a set of coupled nonlinear Schrodinger equations with one equation for each site on the lattice~\cite{heras2024,Real2018}. Each equation includes terms for drive, dissipation and local nonlinearity. We find the steady state solutions using a standard fourth order Runge-Kutta method. Further details are give in Supplementary Discussion 1. The coupling constants between sites were set to represent the Lieb configuration with nearest neighbor hopping. The numerical values for the coupling constants were taken from the experimental PL spectrum (see Fig.~\ref{fig:schematic}c). The pump laser polarisation was linear (horizontally) polarised as in the experiment.

\section*{Data availability}
The data supporting the findings of this study are freely available in the University of Sheffield repository with the identifier \href{https://dx.doi.org/10.15131/shef.data.28360337}{doi:10.15131/shef.data.28360337}.

\section*{Acknowledgments}
This work was supported by the Engineering and Physical Sciences Research Council of the UK, grants number EP/N031776/1 and EP/V026496/1. For the purpose of open access, the author has applied a Creative Commons Attribution (CC BY) licence to any Author Accepted Manuscript version arising.

\section*{Author Contributions}
DNK, PMW, SL and AE designed the experiment. EC grew the planar microcavity wafer by MBE. SL and AE performed experimental measurements with contributions from PMW. PMW performed numerical modelling. SL and PMW analysed the experimental data. PMW led the interpretation of the data and wrote the manuscript with contributions from all authors. We acknowledge the contributions of Deivis Vaitiekus and Ben Royall, University of Sheffield, who etched the lattice structure.

\section*{Competing Interests}
The authors declare that there are no competing interests.


\bibliography{main}




\end{document}